\documentclass[aps,onecolumn,pre]{revtex4}
\usepackage[utf8x]{inputenc}
\usepackage{graphicx}
\usepackage{latexsym}
\usepackage{amsmath}
\usepackage{amssymb}
\usepackage{psfrag}
\usepackage{bm}

\newcommand{\bee}{\begin{eqnarray}}
\newcommand{\eee}{\end{eqnarray}}
\newcommand{\ba}{\begin{array}}
\newcommand{\ea}{\end{array}}
\newcommand{\bc}{\begin{center}}
\newcommand{\ec}{\end{center}}
\newcommand{\bi}{\begin{itemize}}
\newcommand{\ei}{\end{itemize}}

\begin{document}
\title{Translational and rotational Brownian displacements of colloidal particles of complex shapes}%Brownian motion of microparticles}
\author{Bogdan Cichocki}
%\email{cichocki@fuw.edu.pl} 
\affiliation{Institute of Theoretical Physics, Faculty of Physics, University of Warsaw, Pasteura 5,
  02-093 Warsaw, Poland}

\author{Maria L. Ekiel-Je\.zewska}\thanks{Corresponding author: mekiel@ippt.pan.pl}
%\email{mekiel@ippt.pan.pl}%%
 \affiliation{Institute of Fundamental Technological Research,
             Polish Academy of Sciences, Pawi\'nskiego 5B, 02-106 Warsaw, Poland}

\author{Eligiusz Wajnryb}
%\email{ewajnryb@ippt.pan.pl}
 \affiliation{Institute of Fundamental Technological Research,
             Polish Academy of Sciences, Pawi\'nskiego 5B, 02-106 Warsaw, Poland}

\date{\today}

\begin{abstract}
The exact analytical expressions for the time-dependent cross-correlations of the translational and rotational Brownian displacements of a particle with arbitrary shape were 
derived by us in [J. Chem. Phys. {\bf 142}, 214902 (2015) and {\bf 144}, 076101 (2016)]. They are in this work applied to construct %such a powerful 
a method to analyze Brownian motion of 
a particle of an arbitrary shape, and to extract accurately the self-diffusion matrix from the measurements of the cross-correlations, which in turn allows to gain some information on the particle structure. 
%In this way, we significantly improve the procedure proposed in 
%[Phys.~Rev.~E {\bf 88}, 050301 (R) (2013)] to serve the same goal. %We show that the method presented there are in this note used to 
As an example, we apply our new method to analyze 
the experimental results  of D. J. Kraft et al. 
for the micrometer-sized aggregates of the beads [Phys.~Rev.~E {\bf 88}, 050301 (R) (2013)]. We explicitly demonstrate that our procedure, based on the measurements of the  time-dependent cross-correlations in the whole range of times, allows to determine the self diffusion (or alternatively the friction matrix) with a much higher precision than the method based only on their initial slopes. 
%of the measured time-dependent cross-correlations. 
%We also point out that numerical simulations %(similar to those performed in [Phys.~Rev.~E {\bf 88}, 050301 (R) (2013)]) 
%cannot be easily used to extract the self-diffusion matrix with a comparable precision.
Therefore, the analytical time-dependence of the cross-correlations serves as a useful tool to extract information about particle structure from trajectory measurements.
\\

%we demonstrate that 
%there is no need to perform time-dependent numerical simulations similar to those carried out in [Phys. Rev. E {\bf 88}, 050301 (R) (2013) (2013)]. %The only required numerical input is the diffusion (or, equivalently, friction) matrix, which can be evaluated by modeling the particle as an aggregate of beads (even slightly overlapping) and using very precise   {\sc hydromultipole} codes. 
%
%This method is a more efficient and precise alternative to numerical simulations performed in PRE 88, 050301(R) (2013), especially useful for particles with more complex shapes.
%, without any numerical computations, which are time-consuming and have a limited accuracy. %Numerical analysis is not needed.\\
%\bc Abstract \ec
%Brownian motion of a particle with an arbitrary shape is investigated theoretically. Analytical expressions for the time-dependent cross-correlations of the Brownian translational and rotational displacements are derived from the Smoluchowski equation. The role of the particle mobility center is determined and discussed.
\end{abstract}

\maketitle
\section{Introduction}
The characteristic time scales of the translational and rotational Brownian diffusion for nanoparticles are typically much smaller than time resolution of experiments. In this case, nanoparticles can be treated as point-like, % spherical particles, 
and described by the standard Brownian theory \cite{Kampen}. However, for 
microparticles, the characteristic Brownian time scales
are of the order of seconds, and therefore 
non-negligible in comparison to the typical time scales of the measured Brownian motion.
%Nowadays there is a lot of interest in experimental studies of Brownian microparticles \cite{2D,holografic,leptospira,Kraft}.  
For microparticles of complex shapes, a more general theoretical approach is needed to account for the time-dependent Brownian translational and rotational displacements and their cross-correlations. 
Such an approach has been recently developed, and new analytical expressions have been derived from the Smoluchowski equation for 
the Brownian motion of a particle with an arbitrary shape \cite{CEW2012,CEW2015JCP,CEW2016}.

Nowadays there is a lot of interest in experimental studies of Brownian particles of relatively large sizes \cite{2D,holografic,leptospira,Kraft}. Therefore it seems useful to demonstrate explicitly how to apply the theoretical scheme from Refs. \cite{CEW2015JCP,CEW2016} to analyze the data from measurements. In this work, we use the interesting experiment from Ref. \onlinecite{Kraft} as the reference for such a comparison. 
In Ref. \!\onlinecite{Kraft}, the Brownian motions of symmetric and non-symmetric microparticles were investigated at time scales comparable with the characteristic time  of the rotational Brownian diffusion. %, and in this case the standard approach \cite{Kampen} is not sufficient.  Therefore, t
The time-dependent cross-correlations of the Brownian translational and orientational displacements of microparticles with different shapes were measured and the initial slopes of these curves were used to experimentally determine the %mobility and 
friction matrices.   Based on the details of the particle shape, known from the experiment, these matrices were also evaluated numerically with {\sc hydrosub} \cite{hydrosub}, and then used as the input to time-dependent numerical simulations of the Brownian displacement cross-correlations.
Qualitatively, the results agree with each other, but there are significant quantitative differences, which are by the authors explained by the statistical uncertainty of the measurements and irregularities in actual particle shapes. 

In this paper, we look at the results of Ref. \cite{Kraft} from a more general perspective. If similar measurements are performed for a particle of unknown shape, how can the experimental data be used to extract as much details of the particle structure as possible? This information is contained in the mobility matrix (or, equivalently, its inverse called the friction matrix). Therefore, the key point is to construct such a theoretical scheme which allows to determine from the experiments the mobility coefficients (and therefore some information of the particle structure) with the best possible precision. The analytical expressions form Refs. \cite{CEW2015JCP,CEW2016} serve this purpose: they can be used to fit the friction (or, equivalently, mobility) coefficients of a particle using its Brownian displacement cross-correlations in the whole range of the measured times. This procedure allows to determine experimentally the mobility coefficients (and therefore a more detailed structure of the particle) with a significantly higher precision than just taking into account only the initial slope of the correlation functions, as in Ref.~\cite{Kraft}.

\section{Goals and theoretical framework}\label{II}
%\subsection{Goals}
It is worthwhile to consider two generic cases. 

Case 1, analyzed in Sec. \ref{III}. \\
The particle structure and size are known and the goal is to study its translational and rotational Brownian motion. 

Case 2, described in Sec. \ref{IV}. \\
The particle structure is not known, and the Brownian displacement cross-correlations are used to determine its mobility coefficients which in turn provide information about the particle structure and size.

Numerical computations performed in Ref.~\cite{Kraft} correspond to the first case. However, analysis of experimental data is often related to the second case. The challenging question is if the cross-correlations measurements can be analyzed with a precision high enough to provide information about the particle geometry. 

In this paper we apply the analytic expressions from Refs.~\cite{CEW2015JCP,CEW2016} to satisfy both goals. In Sec.~\ref{III}, we perform calculations which belong to the first case. In Sec.~\ref{IV}, we generalize this approach to the second case. 

%....................

The basic theoretical framework used in Sec. \ref{III} is the following.
First, we evaluate   the mobility matrix $\bm{\mu}$, which, by definition, gives the particle translational and rotational velocities when multiplied by the hydrodynamic force and torque exerted by the particles on the fluid \cite{Kim,CEW2015JCP}. We do it for the particles at their initial orientations. 
The inverse of $\bm{\mu}$, called  the friction matrix, is denoted as $\eta \bm{\mathcal H}$, as in Ref. \onlinecite{Kraft},
\bee
\eta \bm{\mathcal H} = \bm{\mu}^{-1}.
\eee
%
%the friction matrix \cite{Kim,CEW2015JCP} for the particles at their initial orientations, to compare with the experimental and numerical results from Ref.~\cite{Kraft}. %studied in Ref.~\cite{Kraft}. 
%We follow Ref.~\cite{Kraft}, where the friction matrix was divided by the fluid viscosity $\eta$ 
%and denoted as $\bm{\mathcal H}$. 
%
Therefore, the translational-translational, rotational-translational and rotational-rotational elements of $\bm{\mathcal H}$ are given in terms of $\mu$m, $\mu$m$^2$ and $\mu$m$^3$, respectively. 

We also evaluate the diffusion tensor $\bm{\mathcal D}$, with all the translational-translational, rotational-translational and rotational-rotational parts,
\bee
\bm{\mathcal D} =\left[\!\! \ba{l} \bm{D}^{tt}  \; \bm{D}^{tr}\\
 \bm{D}^{rt} \;  \bm{D}^{rr} \ea \!\!\right],
 \eee
for the particles at their initial orientations,
\bee
\bm{\mathcal D} = k_BT \bm{\mu},
\eee
where $k_B$ is the Boltzmann constant and  $T$ is the temperature. % and $\bm{\mu}$ is the mobility matrix ($\eta \bm{\mathcal H}$ is the inverse of $\bm{\mu}$). 
%Later on we will use the inverse of $\bm{\mathcal H}$, i.e. the mobility matrix, multiplied by $k_BT$, as the diffusion matrix. 
%

Next, we use the elements of the diffusion matrix $\bm{\mathcal D}$ to determine the cross-correlation matrix ${\mathbf C}(t)$  of the time-dependent Brownian translational and orientational displacements of these particles,\cite{CEW2015JCP,CEW2016}
\bee
{\mathbf C}(t) = \left[\!\!  \ba{c} \left\langle \Delta \mathbf{R}(t)\Delta \mathbf{R}
(t)\right\rangle_0 \; \left\langle \Delta \mathbf{R}%
(t)\Delta \mathbf{u}(t)\right\rangle_0\\ 
\left\langle \Delta \mathbf{u}(t)\Delta \mathbf{R}%
(t)\right\rangle_0 \; \,\left\langle \Delta \mathbf{u}(t)\,\Delta \mathbf{u}%
(t)\right\rangle_0
\ea \!\!\right].\label{C(t)}
\eee 
with $\Delta \mathbf{R}$ and $\Delta\mathbf{u}(t)$ defined as in Refs. \cite{Kraft,CEW2015JCP,CEW2016},
\bee
&&\Delta \mathbf{R}(t) = \mathbf{R}(t) - \mathbf{R}(0),\\
&&
\Delta \mathbf{u}(t)=\frac{1}{2}\sum_{p=1}^{3}\mathbf{u}%
^{(p)}(0)
\times \mathbf{u}^{(p)}(t),\label{defDeltau}
\eee
where $\mathbf{R}(t)$ denotes the time-dependent position of a reference center 
%this mass center 
and $\mathbf{u}%
^{(p)}(t)$, $p=1,2,3$, are 
three mutually perpendicular unit vectors which describe the particle orientation at time $t$ \cite{CEW2015JCP,Kraft}. 

The initial slope of the cross-correlation matrix is related to the diffusion matrix $\bm{\mathcal D}$,
\bee
\left. \frac{d {\mathbf C}(t) }{dt} \right| _{t=0} = 2 \bm{\mathcal D}.
\eee

The averages $\langle ... \rangle_0$ are taken with respect to the particle positions and orientations, using the conditional probability which satisfies the Smoluchowski equation  \cite{Kampen,SE2}.

%\subsection{Method.......}
To determine the hydrodynamic friction matrix $\eta \bm{\mathcal H}$ and the diffusion tensor $\bm{\mathcal D}$ for a given particle (case 1), %as its inverse multiplied by the Boltzmann constant times temperature, 
we solve the Stokes equations, supplemented by the boundary conditions at the particle surface, using the multipole method with the lubrication correction, implemented in the accurate numerical codes {\sc Hydromultipole}  \cite{CEW}. % (the mobility matrix
We apply the multipole truncation order $L\!=\!20$. %, what corresponds to the accuracy ....) 
Then, we apply the expressions for the cross-correlations derived from the Smoluchowski equation in Refs. \cite{CEW2015JCP,CEW2016}.

For spheres or some other symmetric particle shapes, 
the mobility and friction matrices are diagonal. Therefore, the mobility center \cite{Kim} coincides with the center-of-mass, we are in the frame in which the rotational-rotational diffusion tensor is diagonal, % \footnote{For particles of complex shapes, the mobility center can be very far from the center of mass},
and we can directly use the %results
simple analytical expressions 
%for the cross-correlations 
derived 
%from the Smoluchowski 
in Ref.~\cite{CEW2015JCP}. For irregular shapes, we first rotate the system of coordinates to the reference frame in which the rotational-rotational diffusion matrix is diagonal. Still, the translational-rotational coupling does not vanish, and therefore, we use more complicated analytical expressions for the cross-correlations from Ref.~\cite{CEW2016}. To compare with the experiments, we rotate back the frame of reference to the one shown in Fig. 1. 

In Sec.~\ref{III}, this procedure (case 1) will be applied to an experimental example. Sec.~\ref{IV} considers the backwards case 2, where the correlations are
known (experimentally) but the mobility matrix is not.

\section{%Exemplary 
Case 1: Calculations}\label{III}
\subsection{Particles}\label{particles}
\begin{figure}[h!] \psfrag{y}{\small  $\;x_2$} \psfrag{z}{\small $x_3$}
%\includegraphics[height=2.67cm]{reg_trimer.eps} \includegraphics[height=2.67cm]{reg_tetramer.eps}
%\psfrag{y}{\small $\;\;\;x_2$ } \psfrag{z}{\small $x_3$} \includegraphics[height=2.67cm]{irr.eps}\vspace{-0.4cm}
\includegraphics[height=7.2cm]{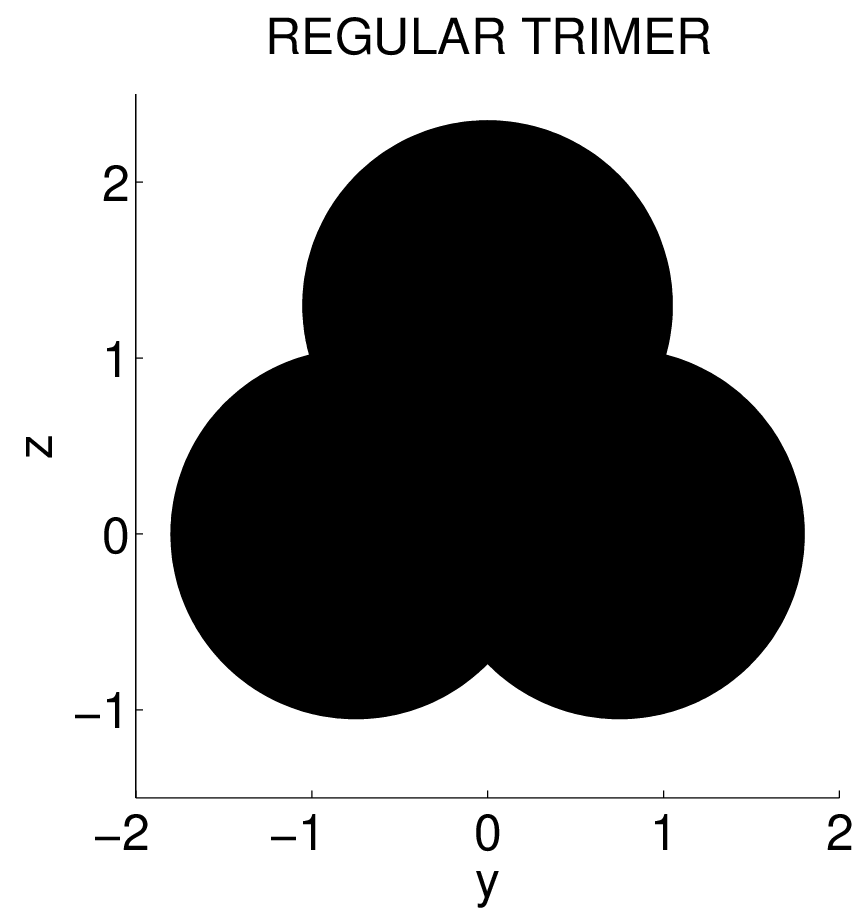}\\

\vspace{0.4cm}
\includegraphics[height=7.2cm]{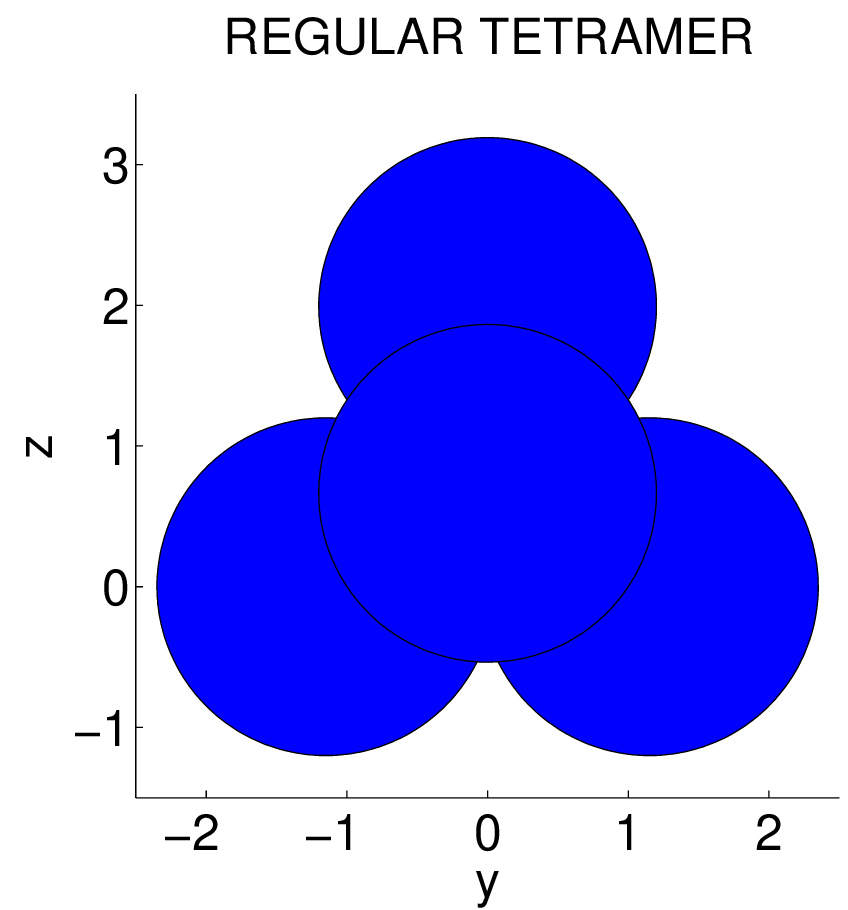}\\

\vspace{0.4cm}
\psfrag{y}{\small $\;\;\;x_2$ } \psfrag{z}{\small $x_3$} \includegraphics[height=7.2cm]{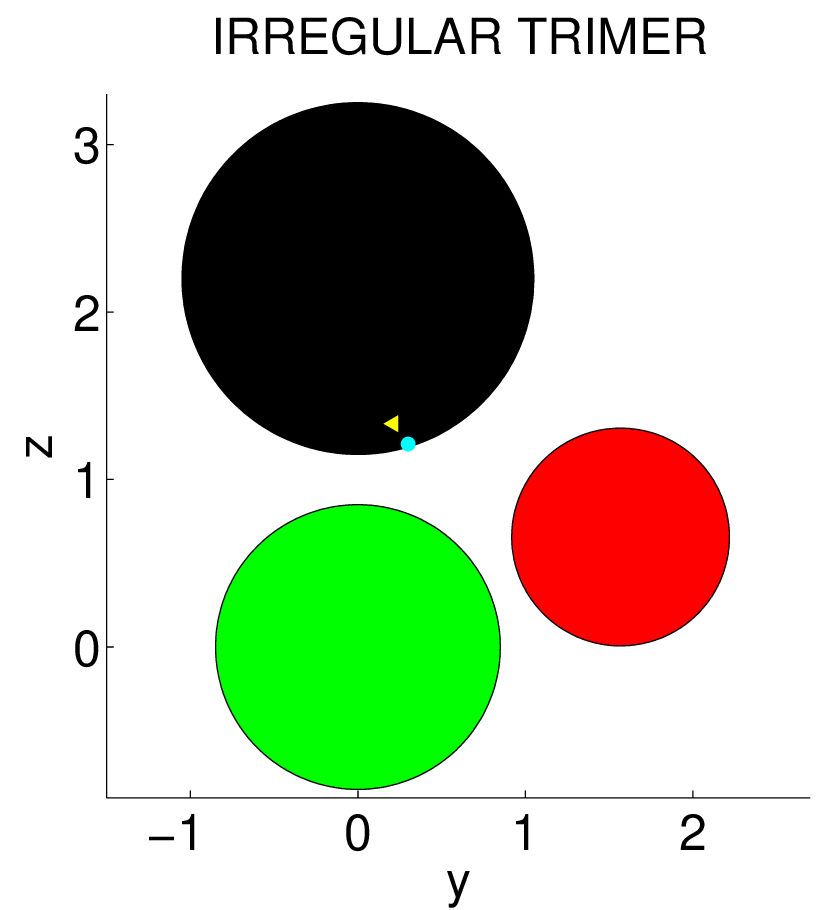}\vspace{-0.4cm}
\caption{Bead models of different rigid particles. For irregular trimer, the center of mass (yellow triangle) does not coincide with the mobility center (cyan circle).}\label{confi} \end{figure}

Following Ref.~\cite{Kraft}, we consider three particles: regular trimer, regular tetramer and irregular trimer, made of spheres (labeled by $i\!=\!1,2,3,4$). 
%, with the following diameters $d_i$ and the center-to-center distances $l_{ij}$. 
For the  
regular trimer and regular tetramer, the beads have equal diameters $d$ and overlap, with equal distances $l$ between the closest bead centers, with $d\!=\!2.1 \;\mu$m and $l\!=\!1.5 \;\mu$m for the trimer and 
%
%(there are large overlaps). 
%For the regular tetramer, diameters of the beads, 
$d\!=\!2.4 \;\mu$m and %, the distance between the closest bead centers, 
$l\!=\!2.3 \;\mu$m for the tetramer. %In both cases, there are overlaps.
For the irregular trimer, the beads have diameters
$d_1\!=\!2.1 \;\mu$m, $d_2\!=\!1.3 \;\mu$m, $d_3\!=\!1.7 \;\mu$m, they do not overlap, and the distances between the bead centers are $l_{13}\!=\!2.2 \;\mu$m, $l_{12}\!=\!2.2 \;\mu$m, $l_{23}\!=\!1.7\; \mu$m. %There are no overlaps.
%\subsection{Friction matrix}
%\subsection{Time scales}
%
%
The particles at their initial orientations with respect to the chosen coordinate system are shown in Fig.~\ref{confi}; the centers of three beads are in the plane $x_1=0$. %, and the particle shapes and orientations with respect to the reference frame are shown in F
%To allow for comparison with the experimental results from Ref.~\cite{Kraft}, 
From now on we will choose the center of mass position as the reference center position $\mathbf{R}(t)$, and will use the same notation as in Ref.~\cite{Kraft}, to allow for the comparison with the experiments.

\subsection{Fluid}
In Ref. \cite{Kraft}, the fluid dynamic viscosity $\eta= 2.22$ mPa s and the temperature $T=294$ K. With these values, time is expressed in seconds and denoted as $t$. 

%\vspace{-0.5cm}
\subsection{Regular trimer}
%\subsection{Friction matrix}
%\subsection{Friction matrix}
\begin{figure*}
\psfrag{t/s}{t}
%\psfrag{C_{ij}(t)/$\mu$m$^2$}{$C_{ij}(t)$}
%\psfrag{C_{ij}(t)/$\mum$}{$C_{ij}(t)$}
%\psfrag{/$\mu m$}{}
%\psfrag{/}{}
%\psfrag{$/\mu m$}{}
%\psfrag{$/\mu$}{}
\psfrag{C_{ii}(t)/\mum^2}{$C_{ii}(t)$}
\psfrag{/\mum}{}

\bc
REGULAR TRIMER\\
\includegraphics[height=3.6cm]{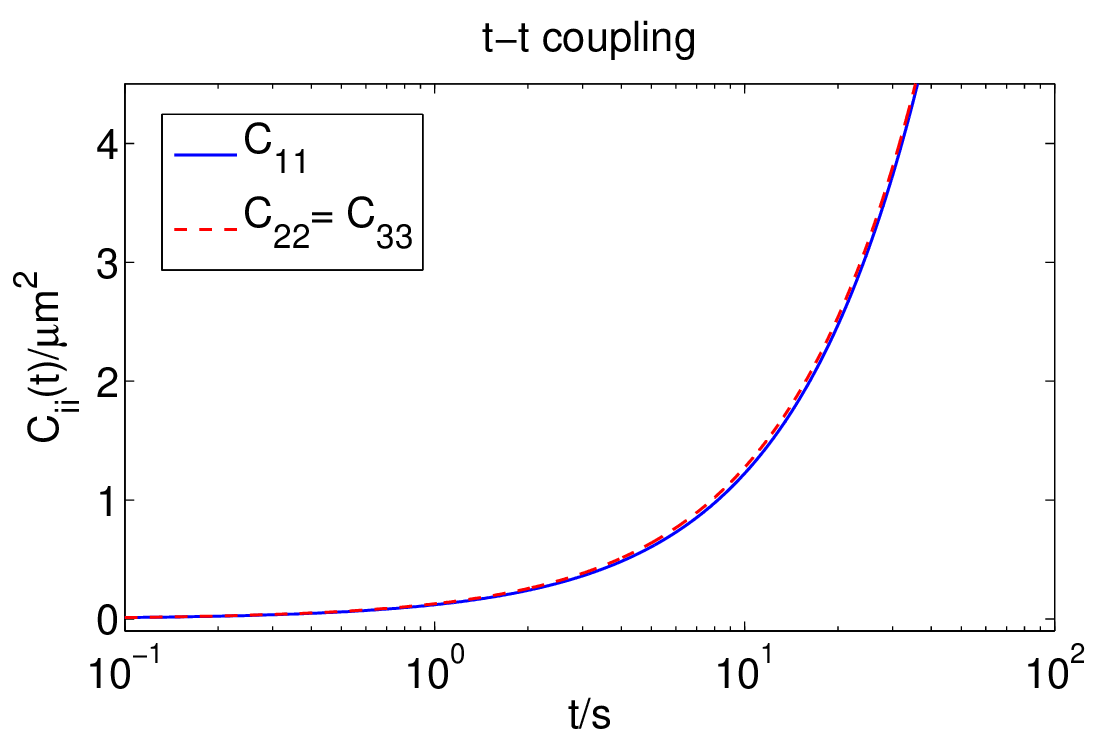} \!\!\!\hspace{0.835cm}REGULAR TETRAMER\hspace{0.835cm}
\!\!\! \includegraphics[height=3.6cm]{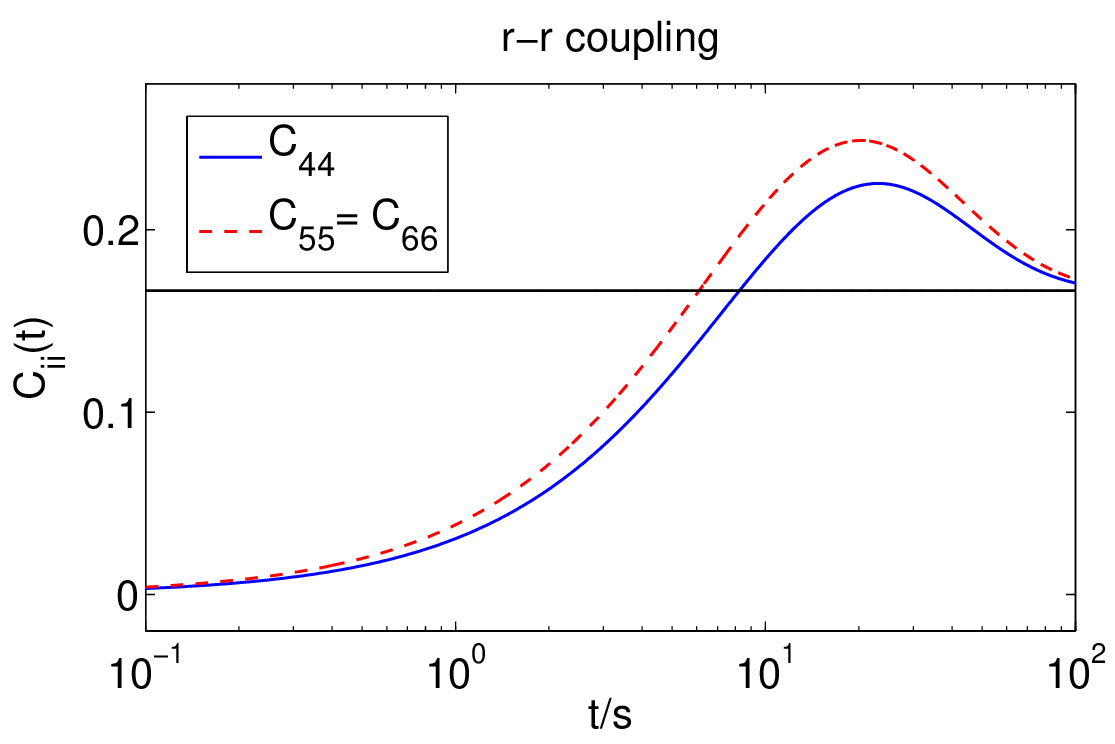}\\
  \includegraphics[height=3.6cm]{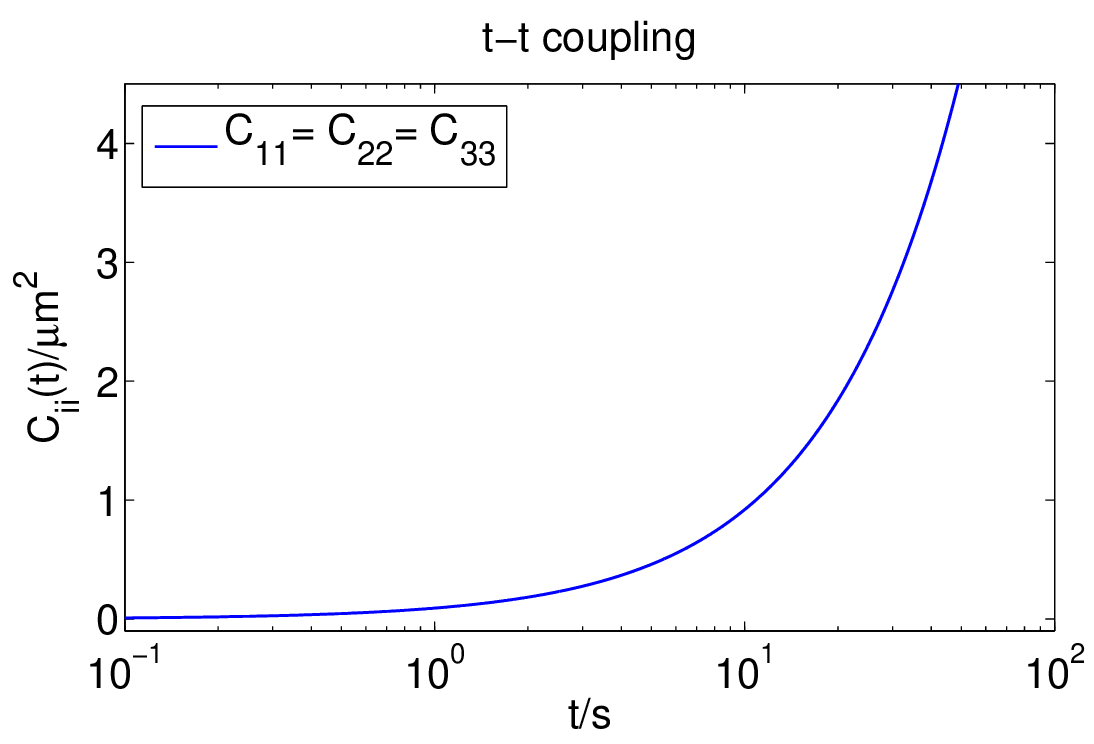} \!\!\!\hspace{1.03cm}IRREGULAR TRIMER\hspace{1.03cm}%\hspace{1cm} REGULAR TETRAMER \hspace{1cm} %\includegraphics[height=5cm]{tetramer_reg_tt.eps} 
  \!\!\!\includegraphics[height=3.6cm]{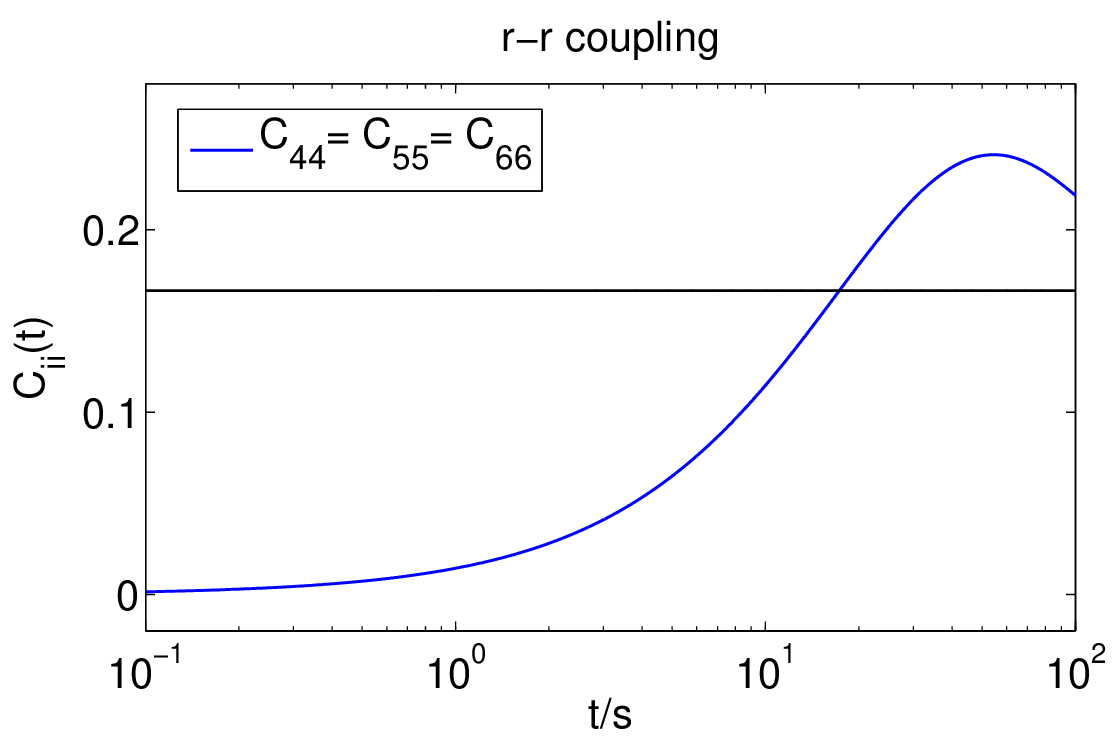} \\
  \vspace{0.3cm}
\! \hspace{-0.1cm}\includegraphics[height=3.6cm]{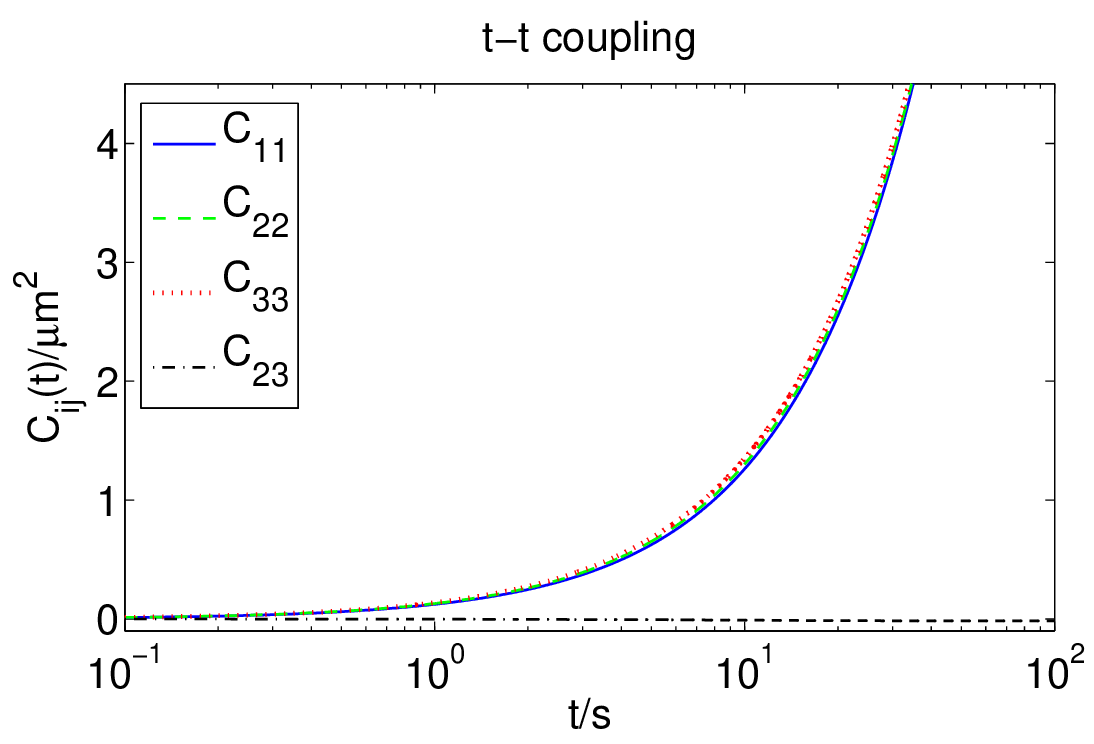}   \!\!\!\!\!
 \includegraphics[height=3.6cm]{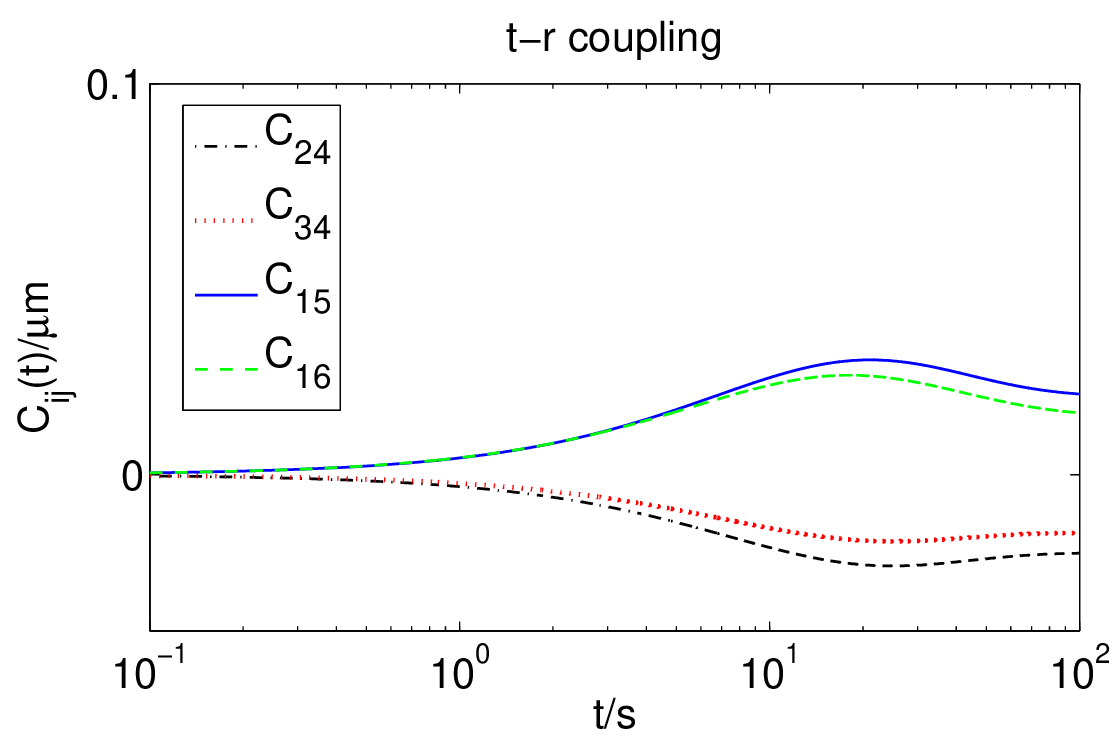}\!\!\!
  \includegraphics[height=3.6cm]{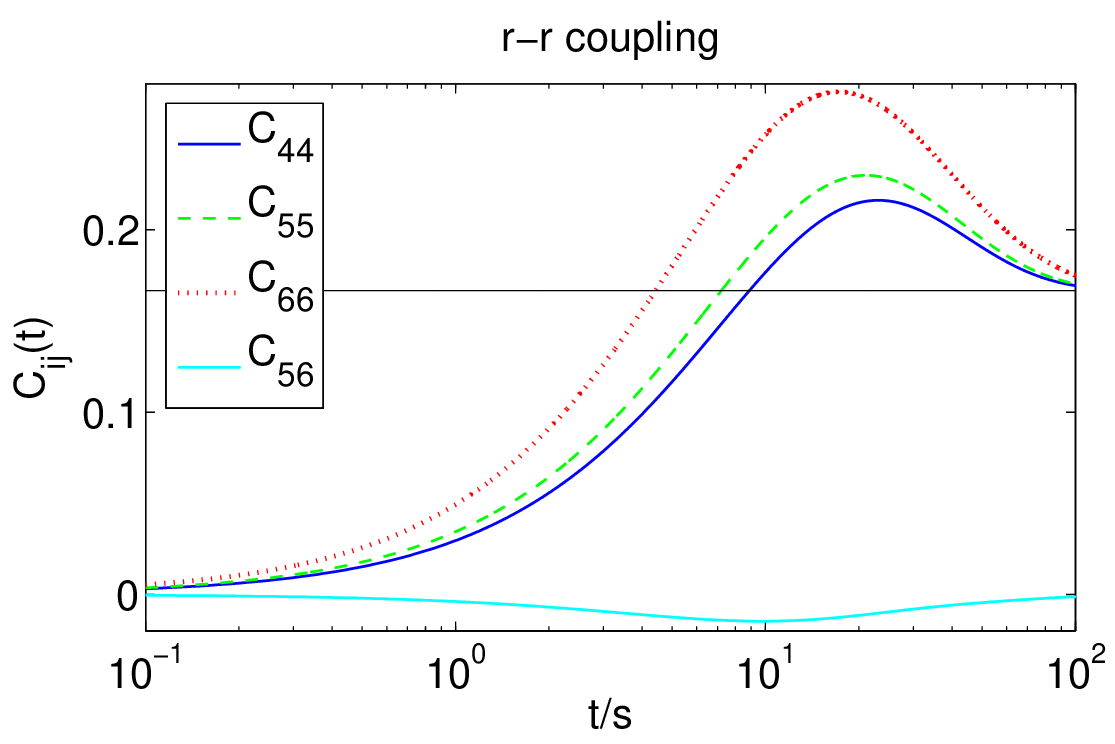}
 \ec
 \vspace{-0.6cm}
 \caption{Cross-correlations $C_{ij}(t)$ for regular trimer (top row), regular tetramer (middle row) and irregular trimer (bottom row). Translational-translational, translational-rotational and rotational-rotational couplings are shown in the left, middle and right columns, respectively. Horizontal line: %the long-time limit 
 $C_{44}(\infty)=C_{55}(\infty)=C_{66}(\infty)\!=\!1/6.$ Here, $t$ denotes time in seconds.}\label{panels}
\end{figure*}

\vspace{-0.4cm}
For the regular trimer at the chosen orientation shown in Fig.~\ref{confi}, the friction matrix has the form,
\bee
\bm{\mathcal H} = \left( \ba{cccccc} 
30.5 & 0 & 0 & 0 & 0 & 0 \\
0 & 28.4 & 0 & 0 & 0 & 0\\
0 & 0 & 28.4 & 0 & 0 & 0\\
0 & 0 & 0 & 112 & 0 & 0\\
0 & 0 & 0 & 0 & 89.3 & 0 \\
0 & 0 & 0 & 0 & 0 & 89.3
 \ea \right),\label{rtrimer}
\eee
%\newpage
with the units of ${\cal H}_{ii}$ equal to 
$\mu$m for $i=1,2,3$ and $\mu$m$^3$ for $i=3,4,5$.

%\subsection{Time scales}

%\subsection{Cross-correlations}
The cross-correlations are given by the same %simple and exact 
expressions as for axially symmetric shapes~\cite{CEW2015JCP}, %. For the experimental system from Ref. \cite{Kraft}, %they have the form,
%Simple analytical expressions, which are exact
%
%
\bee \hspace{-0.4cm}
C_{11}\!&\!\!\!\!=\!\!\!& 0.126\,t-0.0481(1\!-\!e^{-0.123\,t}),\label{rtr11}\\
C_{22}\!&\!\!\!\!\!=\!\!\!\!&C_{33}=\! 0.126\,t+0.0240(1\!-\!e^{-0.123\,t}),\;\;\;\;\\
%\;\;\nonumber\\
%\eee
%
%\bee
C_{44}\!&\!\!\!=\!\!\!& \!\! \frac{1}{6}\!+\!\frac{1}{12}e^{-0.123\,t}\!\!-\!\frac{1}{2}e^{-0.106\,t}\!\!+\!\frac{1}{4}e^{-0.0410\,t}\!,           \nonumber           \\
%&\!+\!&\frac{1}{4}e^{-0.0410\,t},
\\
C_{55}\!&\!\!\!=\!\!\!&C_{66}=          \frac{1}{6}-\frac{1}{6}e^{-0.123\,t}
\!-\!\frac{1}{4}e^{-0.119\,t}                         \nonumber\\
&\!+\!&\frac{1}{4}e^{-0.0367\,t},\label{rtr55}
\eee
where $t$ denotes time in seconds, and the units of ${C}_{ii}$ are equal to 
$\mu$m$^2$ for $i=1,2,3$ and are dimensionless for $i=3,4,5$.

The expressions \eqref{rtr11}-\eqref{rtr55} are plotted versus time in seconds in the top row of Fig. \ref{panels}.
The off-diagonal components vanish. 

\vspace{-0.5cm}
\subsection{Regular tetramer}
%\subsection{Friction matrix}
For the regular tetramer at the chosen orientation shown in Fig.~\ref{confi}, %and the units $\mu$m and $\mu$m$^3$, respectively,
%
%the mobility and friction matrices are diagonal. Therefore, the mobility center coincides with the center-of-mass, we are in the frame in which the rotational-rotational diffusion tensor is diagonal, and we can directly use the results derived in Ref.~\cite{CEW2015JCP}. T
the friction matrix has the form,
%\subsection{Friction matrix}
%For the regular tetramer,
\bee
\bm{\mathcal H} = \left( \ba{cccccc} 
39.7 & 0 & 0 & 0 & 0 & 0 \\
0 & 39.7 & 0 & 0 & 0 & 0\\
0 & 0 & 39.7 & 0 & 0 & 0\\
0 & 0 & 0 & 248 & 0 & 0\\
0 & 0 & 0 & 0& 248  & 0 \\
0 & 0 & 0 & 0& 0 & 248 
 \ea \right),\label{rtetramer}
\eee
with the units of ${\cal H}_{ii}$ equal to 
$\mu$m for $i=1,2,3$ and $\mu$m$^3$ for $i=3,4,5$.

%\newpage
%\subsection{Time scales}

%\subsection{Cross-correlations}
%For the regular tetramer, t
The cross-correlations are given by the same %simple exact 
expressions as for a spherical particle \cite{CEW2015JCP},
%Simple analytical expressions, which ae exact
\bee
C_{11}&\!\!=\!\!&C_{22}=C_{33}= 0.0920\, t,\label{rte11}\\
C_{44}&\!\!=\!\!&C_{55}=C_{66}=\nonumber\\
\frac{1}{6}&\!-\!&\frac{5}{12}e^{-0.0443\,t}+\frac{1}{4}e^{- 0.0148\,t},\;\;\label{rte55}
\eee
where $t$ denotes time in seconds.
The expressions \eqref{rte11}-\eqref{rte55} are plotted versus time in the middle row of Fig. \ref{panels}.
The off-diagonal components vanish.

\vspace{-0.3cm}
\subsection{Irregular trimer}\label{D}

%\subsection{Friction matrix}

For the irregular trimer at the chosen orientation shown in Fig.~\ref{confi}, 
%and the units $\mu$m and $\mu$m$^3$, respectively, 
the friction matrix has the form,
\bee
\bm{\mathcal H}\! = \!\!\left( \!\!\ba{cccccc} 
29.9 & 0 & 0       &    0  & -4.06 & -2.95\\
0 & 28.0 & 0.318   & 2.92  & 0     & 0\\
0 & 0.318 & 26.6   & 2.07  & 0     & 0\\
    0 &2.92& 2.07          & 117 & 0    & 0\\
-4.06 & 0  & 0             & 0   & 101  & 8.51\\
-2.95 & 0  & 0             & 0   & 8.51 & 69.3
  \ea \!\!\right)\!\!,\;\;\label{itrimer}
\eee
%with the units $\mu$m, $\mu$m$^2$ and $\mu$m$^3$, respectively.
with the units of ${\cal H}_{ij}$ equal to 
$\mu$m for $i,j\!=\!1,2,3$ and $\mu$m$^3$ for $i,j\!=\!3,4,5$ and $\mu$m$^2$ for the t-r and r-t
%translational-rotational and rotational-translational 
coefficients.

%\subsection{Cross-correlations}
To determine the cross-correlations, we first go to the frame of reference where the rotational-rotational part of the mobility matrix is diagonal, and evaluate the correlation matrix $\mathbf{C}^{diag}$ in this frame, using the explicit expressions from Ref. \cite{CEW2016}. Then, 
with the use of the $3\!\times \!3$ transformation matrix ${\bm T}$,
we transform it to the original frame of reference, 
\bee
C_{n+i,m+l}=T^{-1}_{ij}C^{diag}_{n+j,m+k}T_{lk},\label{transf}
\eee
%\bee
%{\bf T}=\left[\ba{rrr}
% 1.00000 & 0.000 & 0.000\\
% 0.000 & -0.97144 & -0.23728\\
% 0.000 & -0.23728 & 0.97144\ea
%\right].
%\eee
In Eq.~\eqref{transf}, $i,j,k,l\!\!\!=\!\!\!1,2,3$ are the Cartesian components, 
and $n,m\!\!\!=\!\!\!0,3$ label the translational and rotational parts of the correlation matrix. 
The expressions are %complicated and \!
lengthy and therefore not explicitly written in this note. All the non-vanishing %time-dependent 
translational-translational, translational-rotational and rotational-rotational correlations are plotted in the bottom row of Fig.~\ref{panels}. % Figs.~\ref{irr_tt}, \ref{irr_tr} and \ref{irr_rr}. %, respectively. 

%\vspace{-0.3cm}
\subsection{Discussion}
In general, the numerical  friction tensors and cross-correlations of the Brownian displacements from Ref. \cite{Kraft} agree well with our results presented in this paper. %, based on the exact analytical expressions from Ref.~\cite{CEW2016}. 
No wonder, since they are obtained for the same sizes and relative positions of the spherical beads which model the particle shape.
The only meaningful differences are observed the rotational-translational couplings of the irregular trimer which are small and  difficult to be determined numerically. 
Comparing 
%The observed differences between 
the 
corresponding elements of the hydrodynamic friction matrices in Eqs.~\eqref{rtrimer}, \eqref{rtetramer} and \eqref{itrimer}
with those given in Fig. 1 of Ref.~\cite{Kraft}, we need to take into account %can be related to 
different geometries and accuracies of the models. In the HYDROSUB algorithm and numerical program, used in Ref.~\cite{Kraft},
the surface of the particle is represented by a shell of small elements
(“minibeads”); the results are extrapolated to a zero minibead radius \cite{hydrosub}. In this work, each sphere of the cluster is represented by a single bead, and the accurate HYDROMULTIPOLE numerical codes based on a very precise multipole method corrected for lubrication 
are used to evaluate the friction matrix elements \cite{CEW}. 
%The results in Ref.~\cite{Kraft} indicate that it is difficult to measure accurately the rotational-translational friction coefficients. 

%The cross-correlations for the irregular trimer, evaluated numerically in Ref.~\cite{Kraft}, agree well with the corresponding values determined 
%with the use of the exact analytical expressions from Ref.~\cite{CEW2016} and plotted in the bottom row of Fig.~\ref{panels}, except % The only significant difference appears in 
%the rotational-translational cross correlations which are small and  difficult to be determined numerically. % and experimentally. 
%Our approach allows to determine these (and all the other) correlations with a high accuracy, owing to exact analytical expressions and very precise multipole method corrected for lubrication to determine values of the diffusion matrix.

\section{Case 2: new method}\label{IV} \vspace{-0.4cm}

We will now %consider case 2. We 
apply our analytical expressions from Ref. \cite{CEW2016} to analyze the experimental results  for the irregular trimer given in Ref.~\cite{Kraft}. For this particle,  the non-linear deviations present in the analytical expressions from  Sec.~\ref{D}  are very small, %The first observation is that the experimental 
and the theoretical translational-translational correlations %(and the mean square displacement) 
grow with time almost linearly. We will now estimate the corresponding self-diffusion constant  which characterizes the isotropic mean square displacement at large times. According to the results of Ref.~\cite{CEW2012}, for times much longer that the characteristic scales of the rotational self-diffusion, the mean square displacement is a linear function of time, with the slope which does not depend on the choice of a reference point, and is equal to $6D_{cm}$, where $D_{cm}$ is the translational self-diffusion coefficient for the center of mobility. The center of mobility is such a reference point for which the translational-rotational mobility matrix is symmetric \cite{Kim}. The explicit expression for $D_{cm}$ (see Eq. (20) in Ref.~\cite{CEW2016}) reads,
\bee
\hspace{-0.25cm}&&D_{cm}=\frac{1}{3}\left[ \text{Tr}\bm{D}^{tt}-\sum\limits_{\alpha =1}^{3}\frac{%
(D_{\beta \gamma }^{rt}-D_{\gamma \beta }^{rt})^{2}}{D_{\beta }+D_{\gamma }}%
\right]\!,\;\;\;\;
\eee
 where $(\alpha,\beta,\gamma)$ is a permutation of $(1,2,3)$, $\text{Tr}$ stands for the trace operation, and $D_{\mu}$ are the rotational-rotational diffusion coefficients, defined in
 the frame of reference in which $\bm{D}^{rr}$ is diagonal, i.e. 
 \bee
D^{rr}_{\mu \nu}=D_{\mu} \delta_{\mu \nu},
\eee
with 
$\mu, \nu=1,2,3$.

Using the theoretical and experimental friction matrices, we obtain $D_{cm}=0.065 \; \mu m^2/s$ and $0.073 \;\mu  m^2/s$, respectively. However, the experimental results shown in Fig. 2 in Ref. \cite{Kraft} correspond to value of $D_{cm}$ which is around two times smaller. There is a clear mismatch between the values which characterize essential feature of the Brownian motion: the value deduced from the measured time-dependent translational-translational correlations in a wide range of times and the value determined from the initial slopes of these functions. This difference can be understood taking into account large statistical uncertainty of the experimental results. However, there is no doubt that a method based on fitting $C(t)$ in the whole range of times is more accurate than the method based on the initial slope. 

Therefore, we propose the following new method to determine the self-diffusion matrix $\bm{\mathcal D}$ from the measured time-dependent cross-correlation matrix $\bf{C}(t)$. First, we determine (and go to) the frame of reference in which the matrix $\,\left\langle \Delta \mathbf{u}(t)\,\Delta \mathbf{u}%
(t)\right\rangle_0$ is diagonal. Then, we use the time-dependent analytic expressions from Ref. \cite{CEW2016} to determine in this frame $D_1,\;D_2,\;D_3$, and the rest of the self-diffusion coefficients. Finally, we can transform  $\bm{\mathcal D}$  back to the original frame of reference. This procedure can be applied to 
particles of arbitrary shapes. 

%{\bf Summary}
%\section{Conclusions}
%In this paper, 

Summarizing, in this paper we have demonstrated the applicability of the new method, using the recent experimental results of Ref. \cite{Kraft} as an illuminating example.  
Even if the particles from Ref. \onlinecite{Kraft}  have only a small
difference between the centre of mobility and centre of mass it is worth
further emphasising that the procedure described works for general particles
tracking from an arbitrary tracking point. 

It is very typical for experimentalists to use the initial slope of the cross-correlations of the Brownian translational and rotational displacements (as in Eq. (2) from Ref. \onlinecite{Kraft}) to
determine the diffusion matrix. However, the new method proposed here is based on the full fit of the cross-correlations $C(t)$ in the whole range of time when the measurements have been performed. Therefore, it is
much more accurate.

\acknowledgments
{\small M.L.E.-J. was supported in part by the Polish National Science Centre (Narodowe Centrum Nauki)  under grant No. 2014/15/B/ST8/04359. }

\end{document}